\documentstyle[12pt,epsf]{article}
\def\fo{\hbox{{1}\kern-.25em\hbox{l}}}

\def\beq{\begin{equation}}
\def\eeq{\end{equation}}
\def\eq{\end{equation}}
\def\to{\rightarrow}

\def\bsg{\ifmmode B\to X_s\gamma\else $B\to X_s\gamma$\fi}
\def\bsll{\ifmmode B\to X_s\ell^+\ell^-\else $B\to X_s\ell^+\ell^-$\fi}
\def\bstt{\ifmmode B\to X_s\tau^+\tau^-\else $B\to X_s\tau^+\tau^-$\fi}
\def\shat{\ifmmode \hat{s}\else $\hat{s}$\fi}

\newcommand{\newc}{\newcommand}

\newc{\lcal}{\int {\cal L}dt}
 
\newc{\LSP}{{\chi^0_1}}
\newc{\stauR}{{\tilde \tau_R}}
\newc{\stau}{{\tilde \tau_1}}
\newc{\mstop}{m_{\tilde{t}}}
\newc{\mHpm}{m_{H^\pm}}
\newc{\gsim}{\lower.7ex\hbox{$\;\stackrel{\textstyle>}{\sim}\;$}}
\newc{\lsim}{\lower.7ex\hbox{$\;\stackrel{\textstyle<}{\sim}\;$}}
\newc{\ie}{{\it i.e.}}          
\newc{\etal}{{\it et al.}}
\newc{\eg}{{\it e.g.}}          
\newc{\kev}{\hbox{\rm\,keV}}            
\newc{\mev}{\hbox{\rm\,MeV}}            
\newc{\gev}{\hbox{\rm\,GeV}}            
\newc{\tev}{\hbox{\rm\,TeV}}
\newc{\xpb}{\hbox{\rm\, pb}}
\newc{\xfb}{\hbox{\rm\, fb}}

\newc{\mtop}{m_t}
\newc{\mbot}{m_b}
\newc{\mz}{m_Z}
\newc{\mw}{M_W}
\newc{\alphasmz}{\alpha_s(m_Z^2)}
\newc{\swsq}{\sin^2\theta_W}
\newc{\tw}{\tan\theta_W}
\newc{\cw}{\cos\theta_W}
\newc{\sw}{\sin\theta_W}
\newc{\BR}{\hbox{\rm BR}}
\newc{\zbb}{Z\to b\bar}
\newc{\Gb}{\Gamma (Z\to b\bar b)}
\newc{\Gh}{\Gamma (Z\to \hbox{\rm hadrons})}
\newc{\rbsm}{R_b^\hbox{\rm sm}}
\newc{\rbsusy}{R_b^\hbox{\rm susy}}
\newc{\drb}{\delta R_b}

\newc{\sgn}{\mbox{sgn}}
\newc{\tbeta}{\tan\beta}
\newc{\uL}{{\tilde u_L}}
\newc{\uR}{{\tilde u_R}}
\newc{\cL}{{\tilde c_L}}
\newc{\cR}{{\tilde c_R}}
\newc{\tL}{{\tilde t_L}}
\newc{\tR}{{\tilde t_R}}
\newc{\dL}{{\tilde d_L}}
\newc{\dR}{{\tilde d_R}}
\newc{\sL}{{\tilde s_L}}
\newc{\sR}{{\tilde s_R}}
\newc{\bL}{{\tilde b_L}}
\newc{\bR}{{\tilde b_R}}
\newc{\eL}{{\tilde e_L}}
\newc{\eR}{{\tilde e_R}}
\newc{\mhp}{m_{H^\pm}}
\newc{\mhalf}{m_{1/2}}
\newc{\emt}{{e/\mu /\tau}}

\newc{\lR}{\tilde{l}_R}
\newc{\lL}{\tilde{l}_L}
\newc{\nL}{\tilde{\nu}_L}
\newc{\na}{\chi^0_1}
\newc{\nb}{\chi^0_2}
\newc{\nc}{\chi^0_3}
\newc{\nd}{\chi^0_4}
\newc{\ca}{\chi^{\pm}_1}
\newc{\cb}{\chi^{\pm}_2}
\newc{\camp}{\chi^\mp_1}
\newc{\cbmp}{\chi^\mp_1}
\newc{\capos}{\chi^{+}_1}
\newc{\caneg}{\chi^{-}_1}
\newc{\phit}{\phi_t}
\newc{\phib}{\phi_b}
\newc{\phiew}{\phi_{ew}}
\newc{\htz}{h^0_t}
\newc{\hbz}{h^0_b}
\newc{\hewz}{h^0_{ew}}
\newc{\hsmz}{h^0_{sm}}
\newc{\huz}{h^0_u}
\newc{\hsusyz}{h^0_{susy}}

\def\PLB#1#2#3{Phys. Lett. B {\bf #1}, #3 (19#2)}

\def\PRD#1#2#3{Phys. Rev. D {\bf #1}, #3 (19#2)}
\def\PRL#1#2#3{Phys. Rev. Lett. {\bf#1}, #3 (19#2)}

\def\beq{\begin{equation}}
\def\eeq{\end{equation}}
\def\bea{\begin{eqnarray}}
\def\eea{\end{eqnarray}}
\def\slashchar#1{\setbox0=\hbox{$#1$}           
   \dimen0=\wd0                                 
   \setbox1=\hbox{/} \dimen1=\wd1               
   \ifdim\dimen0>\dimen1                        
      \rlap{\hbox to \dimen0{\hfil/\hfil}}      
      #1                                        
   \else                                        
      \rlap{\hbox to \dimen1{\hfil$#1$\hfil}}   
      /                                         
   \fi}                                         %
\catcode`@=11
\long\def\@caption#1[#2]#3{\par\addcontentsline{\csname
  ext@#1\endcsname}{#1}{\protect\numberline{\csname
  the#1\endcsname}{\ignorespaces #2}}\begingroup
    \small
    \@parboxrestore
    \@makecaption{\csname fnum@#1\endcsname}{\ignorespaces #3}\par
  \endgroup}
\catcode`@=12

\begin{document}

\begin{titlepage}

\null
\vspace*{-3.0cm} 

\begin{flushright}
CERN-TH/99-19 \\
hep-ph/9902242\\
\end{flushright}

\vspace*{20.3cm}

\begin{flushleft}
CERN-TH/99-19 \\
January 1999
\end{flushleft}

\vspace*{-20.0cm}



\huge
\begin{center}
{\Large\bf
Superheavy Supersymmetry}
\end{center}

\large

\vspace{.15in}
\begin{center}

Sandro Ambrosanio${}^{a,b}$ \ and \ James D.~Wells${}^a$

\vspace{.1in}
{\it ${}^{(a)}$ CERN, Theory Division \\
 CH-1211 Geneva 23, Switzerland \\}
\medskip

{\it ${}^{(b)}$ Deutsches Elektronen-Synchrotron DESY \\
  Notkestra\ss e 85, D-22603 Hamburg, Germany \\}

\end{center}
 
 
\vspace{0.15in}
\begin{abstract}

One way to suppress flavor changing neutral currents or CP violating
processes in supersymmetry is to make at least some of the first two 
generations' scalars superheavy (above $\sim 20\tev$).
We summarize the motivations and challenges, theoretically and
phenomenologically, for superheavy supersymmetry.  
We then argue for more viable alternatives on the superheavy theme and 
are led to models where the heavy spectrum follows a pattern of masses 
similar to what arises from gauge-mediation or with a ``hybrid'' spectrum 
of light and heavy masses based on each particle's transformation under 
a global SU(5).  
In the end, despite the differences between the competing ideas, a 
self-consistent natural theory with superheavy masses seems to prefer 
low-energy supersymmetry breaking with possible correlations among the
light sparticle masses.  
The resulting light gravitino and its coupling to matter could also impact 
the discovery capabilities and analyses of these models at Tevatron Run II. 
In addition, we comment on how the presence of superheavy states may influence
the light spectrum, and how this may help efforts to distinguish between 
theories post-discovery.

\end{abstract}


\begin{center}
(Tevatron Run II Workshop on Supersymmetry and Higgs)
\end{center}

\end{titlepage}

\baselineskip=18pt

\vfill
\eject


In the vast space of all viable physics theories, supersymmetry (SUSY) 
is not a point.  Any theory can be ``supersymmetrized'' almost trivially,
and the infinite array of choices for spontaneous SUSY
breaking just increases the scope of possibilities in the real world.
One thing that appears necessary, if SUSY has anything to
do with nature, is superpartners for the standard model particles
that we already know about: leptons, neutrinos, quarks, and gauge
bosons.  These superpartners must feel SUSY breaking and 
{\it a priori} can have arbitrary masses as a result.

Phenomenologically, the masses cannot be arbitrary.  There are several 
measurements that have been performed that effectively limit what
the SUSY masses can be. First, there are direct limits on
$Z\to {\rm SUSY}$, for example, that essentially require all superpartners 
to be above $m_Z/2$.  Beyond this, collider physics limits become model 
dependent, and it is not easy to state results simply in terms of the mass 
of each particle. Second, comparing softly broken SUSY model 
calculations with flavor changing neutral current (FCNC) 
measurements implies that superpartner masses cannot be light and arbitrary.
And finally, requiring that the $Z$ boson mass not result from a fine-tuned
cancellation of big numbers requires some of the particles masses 
be near $m_Z$ (less than about $1\tev$, say).

Numerous explanations for how the above criteria can be satisfied have
been considered.  Universality of masses, alignment of flavor matrices, 
flavor symmetries, superheavy supersymmetry, etc., have all been incorporated 
to define a more or less phenomenologically viable explanation of a softly 
broken SUSY description of nature.  

In this contribution, we would like to summarize some of the basic collider
physics implications of superheavy supersymmetry (SHS) at the Tevatron.  
Our understanding is that analyses of all the specific processes
that are mentioned here in principle are being pursued within other
subgroups. Therefore, our goal in this submission is to succinctly explain 
what SHS is and how some of the observables being studied within other 
contexts could be crucial to SHS.  We also hope that by enumerating
some of the variations of this approach that this contribution could help 
us anticipate and interpret results after discovery of SUSY, and help 
distinguish between theories.
The idea we are discussing goes under several names including ``decoupling 
supersymmetry'', ``more minimal supersymmetry'', ``effective supersymmetry'',
``superheavy supersymmetry'', etc.  
The core principle~\cite{cohen} is that very heavy superpartners do not 
contribute to low-energy FCNC or CP violating processes and therefore cannot 
cause problems. Furthermore, no fancy symmetries need be postulated to keep 
experimental predictions for them under control.

On the surface, it appears that decoupling superpartners is completely
irrelevant for the Tevatron.  After all, Tevatron phenomenology is 
limited to what the Tevatron can produce.  Superheavy superpartners,
which we define to be above at least $20\tev$, are of course not within
reach of a $2\tev$ collider.  However, not all sparticles need be superheavy
to satisfy constraints.  In fact, the third generation squarks
and sleptons need not be superheavy to stay within the boundaries of
experimental results on FCNC and CP violating phenomena.  
As an all important bonus, the third family squarks and sleptons
are the only ones that contribute significantly at one loop to the Higgs 
potential mass parameters.  By keeping the third generation sfermions light, 
we simultaneously can maintain a ``natural'' and viable lagrangian even after 
quantum corrections are taken into account.

In short, the first-pass description of SHS is to say that, in absence
of any alignment, special symmetry or other mechanism yielding 
flavor-horizontal degeneracy, all particles which are significantly coupled 
to the Higgs states should be light, and the rest heavy. 
The gluino does not by itself contribute to FCNC, nor does it couple directly 
to the Higgs bosons and so it could be heavy or light.  
However, the gauginos usually have a common origin, either in grand unified 
theories (GUTs), theories with gauge-mediated supersymmetry breaking (GMSB), 
or superstring theories, and so it is perhaps more likely that the gluino is 
relatively light with its other gaugino friends, the bino and the wino.
Furthermore, the $H_d$ could be superheavy as well, but that is not as 
relevant for Tevatron phenomenology.
Therefore, we can summarize the ``Basic Superheavy Supersymmetry'' (BSHS)
spectrum:
\begin{description}
\item[Superheavy] ($\gsim 20\tev$): $\tilde Q_{1,2}$, $\tilde u^c_{1,2}$, 
 $\tilde d^c_{1,2}$, $\tilde L_{1,2}$, $\tilde e^c_{1,2}$;

\item[Light] ($\lsim 1\tev$): $\tilde Q_{3}$, $\tilde t^c$, 
$\tilde B$, $\tilde W$, $H_u$, $\mu$ (higgsinos);

\item[Unconstrained] (either light or heavy):  
 $\tilde b^c$, $\tilde L_3$, $\tilde \tau^c$, $\tilde g$, $H_d$. 
 
\end{description}
Specific models of SUSY breaking will put the ``unconstrained'' 
fields in either the ``superheavy'' or ``light'' categories. 

Any question about relative masses within each category above can not be 
answered within this framework.
In fact, that is one of the theoretically pleasing aspect of this approach:
no technical details about the spectrum need be assumed to have a
viable theory. Another nice feature is that the mass pattern for 
the scalar partners across generations is somewhat opposite to that 
of the SM fermions. This might well inspire a profound connection 
between the physics of flavor and SUSY breaking. 
A possible theoretical explanation of such a large mass hierarchy in the 
scalar sector is that it could be a result of new gauge interactions
carried by the first two generations only, and which could be, e.g., 
involved in a dynamical breaking of SUSY. For Tevatron enthusiasts, it 
is a frustrating model, since we do not even know what phenomenology should 
be studied because things will change drastically depending on the relative 
ordering of states in the ``light'' category. 

However, there are several features about the BSHS spectrum which are 
interesting not because of the phenomena that it predicts at the Tevatron,
but rather for what it does not predict.
For example, $\tilde q_{1,2} \tilde g$ and $\tilde q_{1,2}\tilde q'_{1,2}$ 
production is not expected at the Tevatron.  This is a potentially large 
source of events in other scenarios, such as minimal supergravity (mSUGRA), 
but is not present here.  A more predictive feature is the expectation of
many bottom quarks and $\tau$ leptons in the final state of SUSY
production.  For example, $p\bar p \to \tilde \chi_2^0 \tilde \chi^\pm_1$ 
will not be allowed to cascade decay through $\tilde e_L$
for example, but may have hundred percent branching fractions to
$\tau$ final states. Therefore, while the ``golden tri-lepton'' signals
are generally suppressed in these models, efforts to look for specific 
$3\tau$ final states are relatively more important to study in the context 
of SHS compared to other models.
Furthermore, light $\tilde t$ and $\tilde b$ production either directly
or from gluino (chargino, stop) decays is of added interest in the BSHS 
spectrum, and may lead to high multiplicity $b$-jet final states.  
In short, drawing production and decay diagrams for all possible permutations 
of the BSHS spectrum always yields high multiplicity $\tau$ or $b$-jet final 
states.  From the BSHS perspective, preparation and analysis for $\tau$ and 
$b$-jet identification is of primary importance. 
For instance, while detection of selectrons and smuons would exclude BSHS, 
detection of many staus and no $\tilde e$ or $\tilde \mu$ would be a good 
hint for it (although one could think of other SUSY scenarios where the 
$m_{\tilde e}-m_{\tilde{\tau}}$ splitting is rather large, due e.g. to 
large values of $\tan\beta$). An interesting place to look for violations
of $e-\tau$ universality is $\chi_1^{\pm}$ or $\chi^0_2$ branching fractions,
after gaugino-pair 
($\tilde \chi_1^+ \tilde \chi_1^-$ or $\tilde\chi_1^{\pm}\tilde\chi_2^0$) 
production.

There are two main problems with the BSHS spectrum. The heavy
particles can generate a disastrously large hypercharge Fayet-Iliopoulos
term proportional to $g^2_1$Tr($Ym^2$).  In universal scalar mass scenarios
these terms are proportional to Tr($Y$) which is zero because of the
gravity--gravity--U(1)$_Y$ anomaly cancellation.  In minimal GMSB scenarios
$m^2\propto Y^2+\cdots$, and so Tr($Ym^2$) = Tr($Y^3$) + $\cdots$
vanishes because of the U(1)$_Y^3$ and SU(N)--SU(N)--U(1)$_Y$ anomaly
cancellation. No such principle exists in the BSHS ansatz given above,
and so the Tr($Ym^2$) is generically a problem.  Barring the possibility
of miraculous cancellations, we can cure the ``Tr($Ym^2$) problem'' by
postulating that the superheavy masses follow a GMSB hierarchy, or that the 
superheavy states come in complete multiplets of SU(5), and the masses of 
all states within an SU(5) representation are degenerate or nearly 
degenerate.  
We will consider both possibilities in the following. 
These requirements may lower the stock of ``superheavy supersymmetry'' ideas 
for some, or it may change how one perceives model building based on 
decoupling superpartners, but it has no direct effect on Tevatron 
phenomenology. 

The superheavy states are inaccessible anyway, so how they arrange their 
masses in detail is of little consequence to us here. On the other hand, 
the generic pattern and theoretical principles beyond this arrangement 
may affect the light sector of the model as well, both directly and indirectly
through higher-order mass corrections. Indeed, another more serious problem, 
which has direct consequence to Tevatron phenomenology is related to 
new two-loop logarithmic contributions to the light scalar masses in
SHS~\cite{arkani}.
For example, the relevant renormalization group equation has a term
\beq
\label{two-loop}
\frac{d \tilde m^2_{\rm light,f}}{d\ln Q} \propto \sum_i \alpha^2_i C^f_i 
\tilde m^2_{\rm heavy} + \cdots
\eeq
where $C^f_i$ are Casimirs for $f$, $i$ labels the indices of the 
SM gauge groups, and $m^2_{\rm heavy}$ is the characteristic superheavy 
mass scale. 
This renormalization group equation begins its running at the scale
where SUSY breaking is communicated to the superpartners.
In supergravity, this is the Planck scale, and so the shift in light
superpartner masses is proportional to the right side of eq.~\ref{two-loop}
multiplied by a large logarithm, of order $\ln M_{\rm Planck}/m_{Z}$. 
This term is so large that in order to keep, e.g., the top squark mass squared
from going negative, it must have a mass greater than several TeV at the high 
scale~\cite{arkani}. 
(Similar problems occur for the other ``light scalars'' which could potentially
put us in a charge or color breaking vacuum.)
Even though the top squark mass can be tuned to be light at the $Z$ scale, the 
renormalization group effects of the heavy top-squark at the high scale
feed into the Higgs sector and results in a fine-tuned Higgs potential.
Since fine-tuning is a somewhat subjective criteria, this problem 
may not be fundamental.  

A healing influence on the above two-loop malady is to make the SUSY 
breaking transmission scale much lower than the Planck scale.
This reduces the logarithm and allows for a more natural Higgs potential
without large cancellations. The most successful low-energy SUSY
breaking idea is GMSB~\cite{giudice}.  
There, the relevant scale is not tied to gravity ($M_{\rm Planck}$),
but rather to the scale of dynamical 
SUSY breaking. Transmission of this breaking to superpartner masses can take
place at scales as low as $\sim m_{\rm heavy}$ in this scheme.  

With some thought about the BSHS spectrum and the troubles that could
arise theoretically from it, we seem to be converging on something
that looks more or less like GMSB. In fact, we can think of the input 
parameters for our converging model to be the input parameters of minimal 
GMSB~\cite{giudice}, which are
\beq
\label{gmsb}
\Lambda, M, N_{\rm mess}, {\rm sign}(\mu), \tan\beta,~{\rm and}~\sqrt{F_0}
\eeq
where $\Lambda$ sets the overall mass scale of the superpartners, 
$M$ is the messenger scale, $N_{\rm mess}$ characterizes the number
of equivalent $5+\bar 5$ messenger representations, and $\sqrt{F_0}$ 
determines the interactions of the goldstino with matter.  
Then we add to these parameters, 
\beq
a_{1,2} = \frac{\tilde m^2_{f_{1,2}}(M)}{\tilde m^2_{f_3}(M)}
\eeq
where we define $\tilde m^2_{f_3}(M)$ to be the minimal GMSB 
values of the sfermion masses at the messenger scale excluding D-terms 
($f=\tilde Q$, $\tilde d^c$, $\tilde L$, $\tilde e^c$). 
The two $a_{1,2}$ parameters with the parameters of eq.~\ref{gmsb}
completely specify a gauge-mediated inspired superheavy SUSY  
(GMSS) model. (Another similar parameter might be introduced for 
the Higgs $H_d$ if this is heavy, but this is less relevant to 
Tevatron phenomenology).  
We suggest that analyses can use these input parameters to make
experimental searches and studies of SHS.
Adding some family dependent discrete symmetries on the superpartners and 
messengers would allow such a model to arise in a similar way as ordinary 
gauge-mediated models.
Recall also that in gauge mediation the Tr($Ym^2$) problem can be solved by 
the triple gauge anomaly rather than by the gravity-gauge anomaly
requirement as would be the case if we had heavy sparticles come in
degenerate remnants of $\bar 5$ and/or $10$ representation,
as a result of the presence of an approximate global SU(5) symmetry. 

The psychological disadvantage of this GMSS model is that it is overkill on 
the FCNC problem.  Gauge mediation cures this problem by itself, and there 
might not be strong motivation to further consider mechanisms that suppress 
it. However, gauge mediation does not automatically solve the CP problem, and 
so the heavy first two generations may help ameliorate it to some degree.  
As an aside, the above discussion can be reinterpreted as a powerful 
motivation for GMSB.
We started with no theory principles but rather only experimental constraints
and with some basic reasoning were drawn naturally to gauge mediation.  
However, we know of no compelling theoretical reason why 
$a_{1,2}\neq 1$.  We only know that if the heavy spectrum follows
a minimal gauge-mediated hierarchy, then the ``Tr($Ym^2$)'' problem can be 
solved. (However, it is possible to construct a more complex gauge-mediated
model that does not satisfy Tr$(Ym^2)=0$.) 
Gauge-mediation, of course, is not necessarily the only way to
transmit low-energy SUSY breaking.  From a phenomenological
point of view, one should be open to a more general low-energy
SUSY breaking framework.  

It must be said that in some cases, even when SUSY breaking is 
transmitted at low scales as in GMSS, one 
still could have a hard time avoiding 
color- and charge-breaking vacua.
Indeed, the contribution from the superheavy 
states in eq. \ref{two-loop} can still be large when loops from all the 
scalars of the first two generations add up. 

As anticipated, another possibility to cure the ``Tr($Ym^2$) problem'' and 
the ``two-loop problem'' is with the hybrid multi-scale SUSY models 
(HMSSM)~\cite{hybrid}, using the ``approximate global SU(5)'' pattern:

\begin{description}

\item[HMSSM-I:] The first two generations of the $10$ representation
of SU(5) ($\tilde Q_{1,2}$, $\tilde u^c_{1,2}$, 
$\tilde e^c_{1,2}$) are superheavy ($\tilde m_{10_{1,2}}$), while the
rest of the sparticles are light and approximately degenerate. 

\item[HMSSM-II:] In HMSSM-IIa
all three generations of the $\bar 5$ representation of SU(5) 
($\tilde d^c_{1,2,3}$,  $\tilde L_{1,2,3}$) are superheavy 
($\tilde m_{\bar 5_{1,2,3}}$), while the rest of the sparticles are light.  
In HMSSM-IIb just the first two generations of the $\bar 5$ are superheavy 
($\tilde m_{\bar 5_{1,2}}$).

\end{description}

In these models, one attempts a solution of the FCNC problem by using 
a combination of some decoupling (superheavy scalars) and some degeneracy.  
A theoretical motivation for this could be that due to an approximate 
SU(5) global symmetry of the SUSY breaking dynamics, only some of the 
quark/leptons superfields with the same SU(3)$\otimes$SU(2)$\otimes$U(1)
quantum numbers are involved in the SUSY breaking sector, carry an additional
quantum number under a new ``strong'' horizontal gauge group and are 
superheavy. 
The other superfields instead couple only weakly (but in a flavor-blind way) 
to SUSY breaking and are light and about degenerate.  
 
Actually, these ``hybrid'' models present many advantages compared  
to other SHS realizations. The reduced content of the superheavy 
sector considerably weakens the ``two-loop'' problem, since the negative 
contribution to the light scalar masses squared is less important. This 
is especially true for the HMSSM-II, and in particular the IIb version. 
Actually, it is in this case possible to raise the $m_{\rm heavy}$ scale 
up to $\sim 40\tev$, in a natural way. 
Most problems with FCNC phenomena come from $L-R$ operators, and since
these operators remain suppressed, the hybrid models
are phenomenologically viable and attractive versions of superheavy
supersymmetry.

The resulting spectrum is different than GMSS and BSHS in that some of 
first two generation states are now allowed to be light. 
For example, in the HMSSM-I model, the $\tilde L$ sleptons can be light, 
and on-shell decays of winos into $L+\tilde L$ can allow the trilepton 
signal $\tilde \chi^\pm_1 \tilde\chi^0_2\to 3l$ to have
near $100\%$ branching fraction.  This is not possible in the BSHS spectrum.
Also, it may be useful to study the total rate of jets plus missing energy 
and the kinematics of the events to discern that only $\tilde d^c$, 
$\tilde s^c$ and 3rd-generation squarks are light, and the remaining 
squarks are heavy. More detailed phenomenological studies might start
from observing that in the ``hybrid'' case too, one still needs 
low-energy SUSY breaking to deal with the ``two-loop problem''. 
Again, a GMSB-inspired spectrum for the light sector corrected by 
the (here reduced) presence of the heavy scalars seems relevant as a 
starting point. In this case, a parametrization along the lines described
above for the GMSS would involve new additional parameters such as 
$\tilde m_{10_{1,2}}$ for the HMSSM-I or 
$\tilde m_{\bar{5}_{1,2(,3)}}$ for the 
HMSSM-IIa(,b), plus possibly an analogous parameter for $H_d$.

Whether the spectrum is more minimal GMSB-like or is better 
described by the ``hybrid models'', there is one feature in common.  
Due to the ``two-loop problem'', SHS appears more natural with low-energy 
supersymmetry breaking, independent of how the SUSY breaking 
and transmission are accomplished 
(minimal gauge-mediation ideas or otherwise).  This implies that the lightest 
superpartner is the gravitino rather than the neutralino, as e.g. in mSUGRA. 

Depending on the details of SUSY breaking and the transmission of that 
breaking to superpartners (e.g., whether $\sqrt{F_0}$ in eq. \ref{gmsb} 
is much larger or smaller than about $100\tev$), the next-to-lightest 
superpartner (NLSP) will either decay promptly in the detector, or decay 
with a long lifetime outside the detector.  This may very well dominate the 
phenomenological implications of the model. Another important feature is the 
identity of the NLSP. It is well known that, e.g. in a GMSB-like spectrum, 
the best candidates are the $\tilde \chi^0_1$ and the lightest stau 
$\tilde \tau_1 \simeq \tilde \tau_R$. 
In SHS, a scenario with a neutralino NLSP, with associated decays such as 
$\tilde \chi^0_1\to \gamma \tilde G$ possibly inside the detector, is still 
an important possibility. In this case, multiple high-$p_T$ photons are the 
tags to spectacular events.
On the other hand, in the GMSS model and in the HMSSM models, the 
$\tilde \tau_R$ is always part of the light scalar sector. 
In addition, here the negative contributions from the heavy scalars
to its mass will tend to lower it compared to the neutralino mass. 

Further, in many realizations of SHS the big mass hierarchy between the 
Higgses $H_u$ and $H_d$ can trigger very large ${\cal O}(m_{\rm heavy}/M_Z)$ 
values of $\tan\beta$ (which might provide a reasonable explanation of
the large $m_t/m_b$ ratio without fine-tuning). 
As a side result, after $L-R$ mixing, the $\tilde \tau_1$ mass might 
turn out to be even lighter relative to the other scalars and the 
neutralino than in GMSB models. Hence, we believe that the possibility 
of a $\tilde \tau_1$ NLSP in SHS deserves very serious consideration. 
If this is the case, the NLSP is charged and might live beyond the detector
if $\sqrt{F_0}$ is relatively large.  
Then stable charged particle tracks in the calorimeter will 
be tags to even more spectacular events~\cite{dimo}.  
Many of the results in the
gauge-mediation literature will directly apply for {\it discovery}.
After discovery, the particles that come along with the spectacular
stable charged tracks (SCTs) or the high-$p_T$ photons can then be studied to 
find out with great confidence the light particle content of the theory, 
that could distinguish between the superheavy and the ``traditional''
models. 

As an example of distinguishing phenomenology, we can define 
$R_{\ell' \ell}$ to be
\beq
R_{\ell' \ell} \equiv \sum_{\stackrel{\scriptscriptstyle\ell=e,\mu }
{\scriptscriptstyle\ell'=e,\mu}}
\frac{\sigma[2 \; {\rm SCTs} + \ell'^+ \ell^-]}
{\sigma[2 \; {\rm SCTs} + X]}.
\eeq
From total SUSY production in HMSSM-IIb one expects 
$R_{\ell'\ell} < 1/10$ since $\tilde e$ and $\tilde \mu$ cannot participate
in the decays.  Most events will then have 
$X=\tau^+_{\rm hard}\tau^-_{\rm hard}$ accompanying the 2~SCTs.
However, in minimal GMSB the $\tilde e$ and $\tilde \mu$ are present
in the low-energy spectrum, and so
$\tilde \chi^+\to e^+ \nu_e \tau^\pm_{\rm soft}\tilde \tau^\mp$
may proceed with large branching fraction.  Although 
a precise number depends mainly on the number of
messenger representations and $\tan\beta$,
$R_{\ell'\ell}$ could be greater than $1/2$ in GMSB.
More generally, the unusually large 
$L-R$ mass hierarchies that are typical of ``hybrid'' models may allow
identification of observables suitable for discerning superheavy supersymmetry
from other more conventional forms of supersymmetry at the Tevatron.

{\bf Acknowledgements:} We are indebted to A.~Nelson and S.~Martin for
helpful comments. 


\end{document}